\documentclass{moriond}

\usepackage{bold-extra}
\usepackage{amssymb}
\usepackage{xspace}
\usepackage{graphicx}
\usepackage[export]{adjustbox}
\usepackage{subfig}
\usepackage{placeins}
\usepackage{caption}

\usepackage{tikz}

\newcommand{\swatch}[1]{\tikz[baseline=-0.6ex]\node[fill=#1,shape=rectangle,draw=black,thick,minimum width=5mm,rounded corners=0.5pt](){};}
\newcommand{\met}{\ensuremath{E_T^{\rm miss}}}

\definecolor{powderblue}{HTML}{B0E0E6}
\definecolor{blue}{HTML}{0000FF}
\definecolor{cadetblue}{HTML}{5F9EA0}
\definecolor{navy}{HTML}{000080}
\definecolor{silver}{HTML}{C0C0C0}
\definecolor{wheat}{HTML}{F5DEB3}
\definecolor{snow}{HTML}{FFFAFA}
\definecolor{cornflowerblue}{HTML}{6495ED}
\definecolor{turquoise}{HTML}{40E0D0}
\definecolor{green}{HTML}{008000}
\definecolor{darkorange}{HTML}{FF8C00}
\definecolor{orangered}{HTML}{FF4500}
\definecolor{thistle}{HTML}{D8BFD8}

\newcommand{\MZP}{\ensuremath{M_{Z^{\prime}}}\xspace}
\newcommand{\GZP}{\ensuremath{\Gamma_{Z^{\prime}}}\xspace}
\newcommand{\cotH}{\ensuremath{\cot\theta_\textrm{H}}\xspace}
\newcommand{\ZP}{\ensuremath{Z^{\prime}}\xspace}

\newcommand{\sw}[1]{\textsf{#1}\xspace}

\newcommand*{\POWHEG}{\sw{POWHEG}}

\newcommand*{\RIVET}{\sw{Rivet}}

\newcommand*{\CONTUR}{\sw{Contur}}

\newcommand*{\PBZPWP}{\sw{PBZpWp}}
\newcommand*{\hvq}{\sw{hvq}}
\newcommand*{\HERWIG}{\sw{Herwig}}

\def\Journal#1#2#3#4{{#1} {\bf #2}, #3 (#4)}

\def\be{\begin{equation}}
\def\ee{\end{equation}}
\def\bea{\begin{eqnarray}}
\def\eea{\end{eqnarray}}

\newcommand{\Photo}{}

\begin{document}
\vspace*{4cm}
\title{Exploring Contur beyond its default mode:
a case study}

\author{ M. M. Altakach$^{1}$ \footnote{Speaker}, J. M. Butterworth$^2$,
T. Je\v{z}o$^3$,
M. Klasen$^3$,
I. Schienbein$^4$ }

\address{$^1$Institute of Theoretical Physics, Faculty of Physics, University of Warsaw, \\ ul.~Pasteura 5, PL-02-093 Warsaw, Poland\\
 $^2$ Department of Physics \& Astronomy, UCL, Gower~St., WC1E~6BT, London, UK\\
 $^3$Institut f\"ur Theoretische Physik, Westf\"alische
 Wilhelms-Universit\"at M\"unster, \\ Wilhelm-Klemm-Stra\ss{}e 9, 48149
 M\"unster, Germany \\
 $^4$Laboratoire de Physique Subatomique et de Cosmologie,
 Universit\'e Grenoble-Alpes, CNRS/IN2P3, \\
 53 Avenue des Martyrs, 38026 Grenoble, France}

\maketitle\abstracts{
We discuss \CONTUR's different modes by studying a leptophobic Top-Colour (TC) model. We use, for the first time, higher order calculations for both the signal (NLO) and the background (up to NNLO). 
We compare the results between the different approaches of \CONTUR. Furthermore, we compare these results to the ones coming from a direct search.
}

\section{Introduction}
\label{section-1}

Among the many goals of the Large Hadron Collider (LHC) is to search for physics Beyond the Standard Model (BSM).
However, in its first two runs, no clear signals of New Physics were seen.
Nevertheless, hundreds of measurements were published. Thus, an interesting approach would be to use these already existing measurements to test at what significance the new theoretical ideas are already excluded.
A toolkit that was designed to exactly do this is the Constraints On New Theories Using Rivet (\CONTUR) toolkit~\cite{JMB}~\cite{Buckley:2021neu}. The idea behind \CONTUR is that BSM models implemented in Monte Carlo event generators can be compared to cross section fiducial measurements at particle level in a model independent way. It uses the hundreds of measurements already included in the \RIVET~\cite{rivet} (Robust Independent Validation of Experiment and Theory) repository to constraint the phase space of the BSM model.

\section{Different modes of \CONTUR}
\label{section-2}

The default mode of \CONTUR is used to look for striking deviations from the Standard Model (SM). In this scenario, \CONTUR sets the SM background to the data. It then adds the signal to the data and checks whether there is a room for the BSM signal within the uncertainties of the data. The second, more complete approach, that can be performed by \CONTUR is when the SM model predictions for the measurements that we are interested in are included. In this case, the SM background is set to be equal to the SM predictions. \CONTUR then adds the signal on top of the SM theory prediction and compares the result to the data within uncertainties. Finally, \CONTUR can also be used in the so-called ``expected limit" mode. Here, the central value of the measurement is shifted to lie exactly on the simulated SM prediction while keeping the uncertainties of the measurement. The exclusion limits are then evaluated in the same way as in the previous case. 

\section{Signal and background calculation}

In this case study, the signal comes from a leptophobic TC model~\cite{lepTC}. This model has an additional $SU(3)$ symmetry under which only the third generation transforms. The first and second generations then transform under the original $SU(3)$ symmetry. To prevent the bottom quark from being as massive as the top, an additional $U(1)$ symmetry associated with a \ZP boson is considered. The two $SU(3)$s and $U(1)$s are then broken down to the $SU(3)_C$ and $U(1)_Y$ of the SM, respectively. As suggested by the name of the model, the \ZP does not couple to leptons. Furthermore, it does not couple to the second generation quarks. The two input parameter of this TC model are the mass of \ZP (\MZP) and the ratio of the two $U(1)$ coupling constants (\cotH). However, using the decay width of the \ZP boson (\GZP), one can work with $\GZP/\MZP$ instead of \cotH~\cite{ConturTC}. We perform two calculations of the signal, one with \HERWIG at LO including all the two to two processes, with the \ZP boson either in the s-channel or as an external outgoing leg, and one with the \PBZPWP package at NLO only considering the \ZP to $t\bar{t}$ final state~\cite{altakach:1}.

The partonic $t\bar{t}$ cross section at NLO accuracy reads:
\begin{equation}
\sigma_{ab}(\mu_r) = \sigma_{1;1}(\alpha_S \alpha) + \sigma_{2;0}(\alpha_S^2) + \sigma_{0;2}(\alpha^2) + \sigma_{3;0}(\alpha_S^3) + \sigma_{2;1}(\alpha_S^2 \alpha) + \sigma_{1;2}(\alpha_S \alpha^2) + \sigma_{0;3}(\alpha^3)
\label{eq:background}
\end{equation}
The indices $a$ and $b$ represent the power in $\alpha_S$ and $\alpha_{EW}$, respectively. For our analysis, we simulate the SM QCD $t\bar{t}$ background ($\sigma_{2;0}$) and the NLO QCD correction to it ($\sigma_{3;0}$) using the \hvq package~\cite{hvq}. We also calculate the electroweak (EW) SM $t\bar{t}$ production ($\sigma_{0;2}$), the NLO QCD correction to it ($\sigma_{1;2}$), and the interference between QCD and EW SM top-pair production ($\sigma_{1;1}$) using \PBZPWP. Furthermore, we include, only at 13 TeV, a sample of QCD corrections to $\sigma_{2;0}$ up to NNLO accuracy obtained with the \texttt{ttJ\_MiNNLO} package~\cite{NNLO}. Finally, using the \texttt{dijet} package~\cite{dijet}, we simulate the ATLAS inclusive jet and dijet cross section measurement~\cite{ATLAS} at NLO accuracy.

\section{Results}

\begin{figure}[h!]
  \begin{center}
    \subfloat[]{\includegraphics[width=0.4\textwidth]{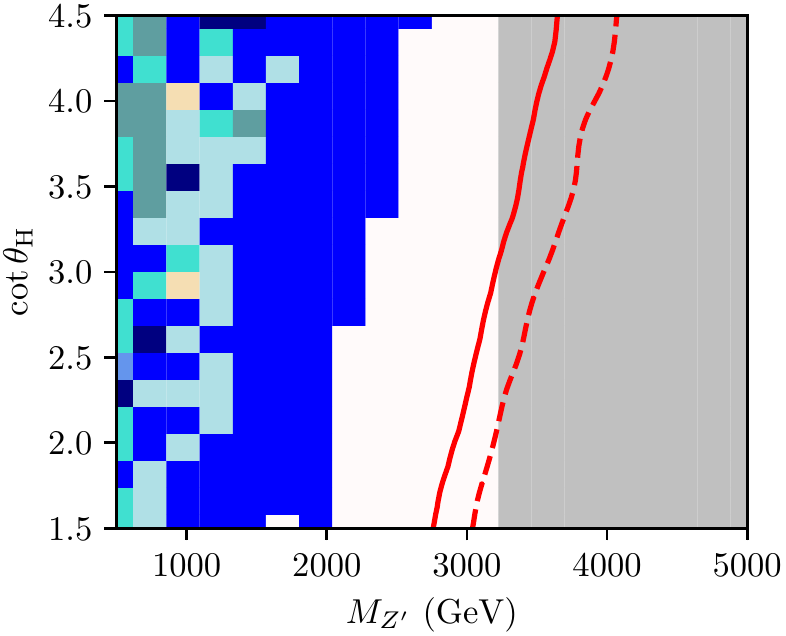}\label{fig:ph_data}}
    \subfloat[]{\includegraphics[width=0.4\textwidth]{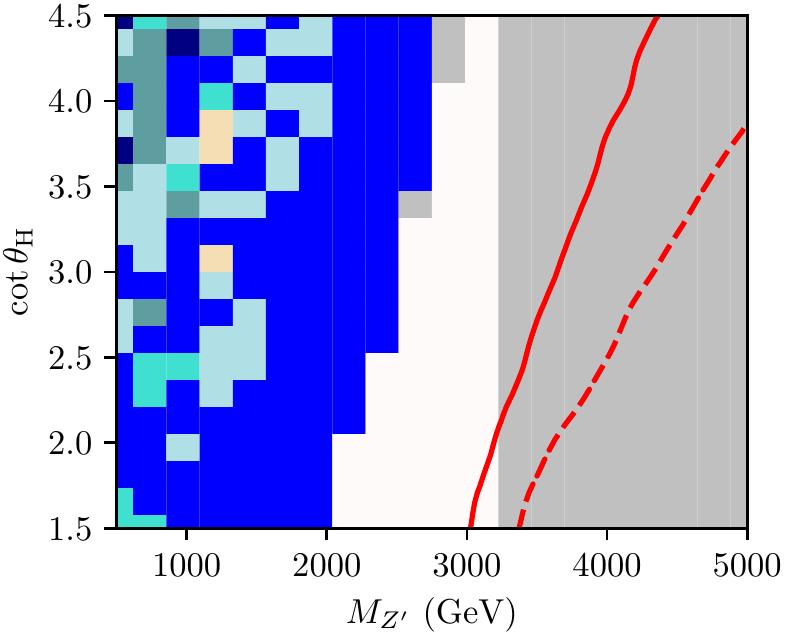}\label{fig:hw_data}}
    \caption{  -- Exclusion limits for the leptophobic TC model in the \MZP versus the $\cot \theta_H$ plane.
    The coloured blocks in the legend shows the breakdown into
the most sensitive analysis pool for each scan point. The 95\% CL (solid red) and 68\% CL exclusion (dashed red) contours are superimposed,
    using the default mode of \CONTUR (background = data). (a): the $t\bar{t}$ signal is obtained at NLO accuracy using the \PBZPWP \POWHEG package. (b): the signal is obtained using \HERWIG (inclusive LO).}
    \label{fig:phhw_data}
    \begin{tabular}{llll}
        \swatch{powderblue}~CMS $\ell$+\met{}+jet &
        \swatch{blue}~ATLAS $\ell$+\met{}+jet &
        \swatch{cadetblue}~ATLAS $e$+\met{}+jet \\
        \swatch{navy}~ATLAS $\mu$+\met{}+jet &
        \swatch{silver}~ATLAS jets &
        \swatch{wheat}~CMS Hadronic $t\bar{t}$ \\
        \swatch{snow}~ATLAS Hadronic $t\bar{t}$ &
        \swatch{cornflowerblue}~ATLAS $\ell_1\ell_2$+\met{} &
        \swatch{turquoise}~ATLAS $\ell_1\ell_2$+\met{}+jet \\
    \end{tabular}
\end{center}
\end{figure}

Starting with the default mode of \CONTUR, we show in Figure~\ref{fig:phhw_data} results using the data as background in the the \MZP versus the $\cot \theta_H$ plane. While we use \PBZPWP to calculate the signal at NLO in Figure~\ref{fig:ph_data}, in Figure~\ref{fig:hw_data} we use \HERWIG to simulate it inclusively at LO. The strongest sensitivity in both cases come from the various top measurements at low masses, the ATLAS fully hadronic measurements for masses between 2 and 3 TeV, and the jet measurements at high masses. It is visible that stronger limits are obtained in the \HERWIG case. This is due to the fact that at LO, not only the $t\bar{t}$ final state is included, but also the $u\bar{u}$ and $d\bar{d}$ final states.

\begin{figure}[h!]
  \begin{center}
    \subfloat{\includegraphics[width=0.4\textwidth]{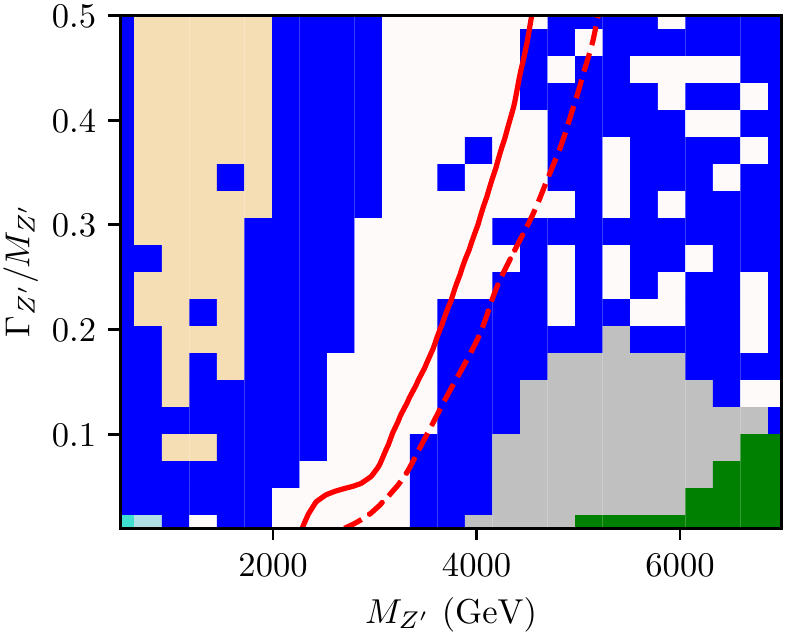}\label{fig:ph_data_to}}
    \subfloat{\includegraphics[width=0.4\textwidth]{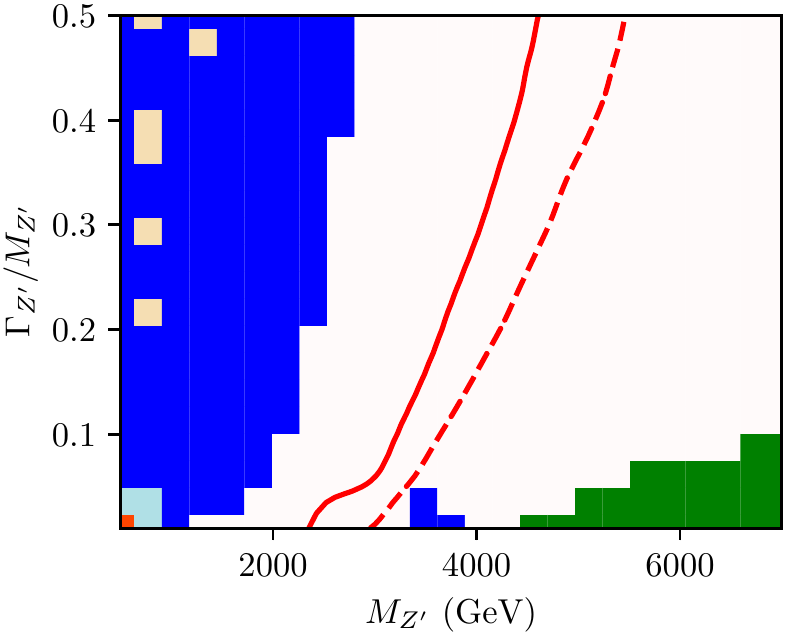}\label{fig:ph_nlo}}\\ [-2ex]
    \subfloat{\includegraphics[width=0.4\textwidth]{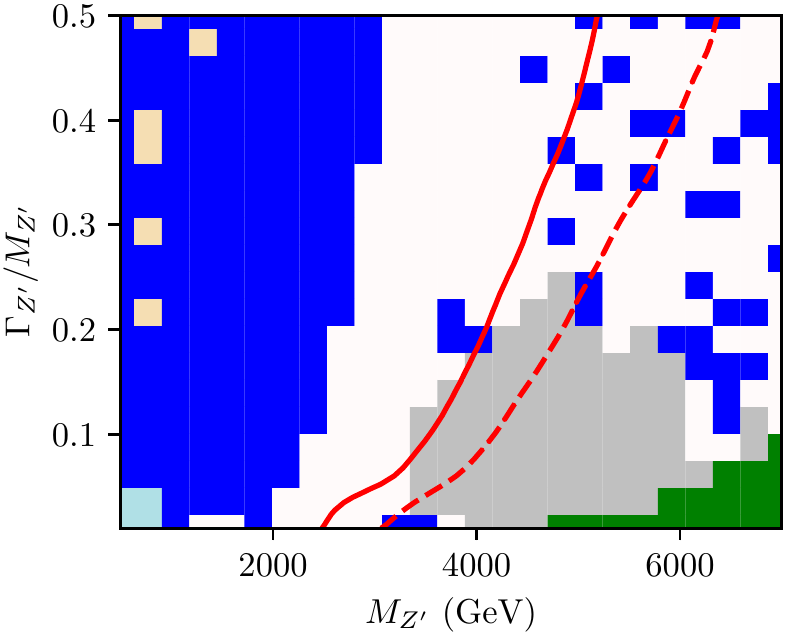}\label{fig:ph_nnlo}} 
    \subfloat{\includegraphics[width=0.4\textwidth]{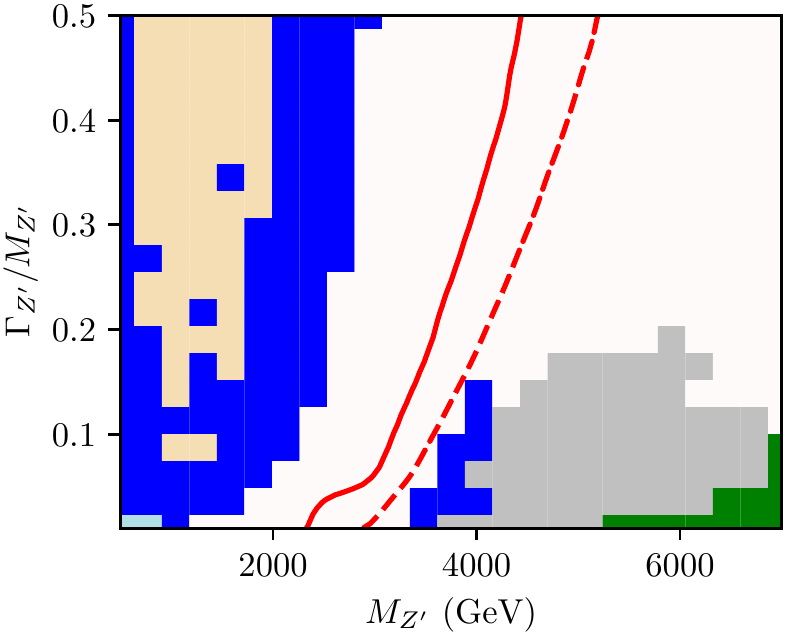}\label{fig:ph_nnlo_exp}} \\
    \caption{  -- Same as Figure~\ref{fig:phhw_data}, but in the \MZP versus $\GZP/\MZP$ plane, for $\ZP \rightarrow t\bar{t}$ (\PBZPWP).
    Top left: default mode, but only using measurements where SM predictions are available.
    Top right: background = NLO SM predictions.
    Bottom left: background = NNLO SM predictions where available, and NLO elsewhere.
    Bottom right: expected limit using NNLO predictions.
}
    \label{fig:ph-smth} 
      \begin{tabular}{llll}
        \swatch{snow}~ATLAS Hadronic $t\bar{t}$ &
        \swatch{wheat}~CMS Hadronic $t\bar{t}$ &
        \swatch{green}~ATLAS \met{}+jet \\
        \swatch{blue}~ATLAS $\ell$+\met{}+jet &
        \swatch{silver}~ATLAS jets &
        \swatch{powderblue}~CMS $\ell$+\met{}+jet \\
        \swatch{darkorange}~ATLAS $\mu\mu$+jet &
        \swatch{orangered}~ATLAS $ee$+jet &
        \swatch{turquoise}~ATLAS $\ell_1\ell_2$+\met{}+jet 
\end{tabular}  
\end{center}         
\end{figure}

In Figure~\ref{fig:ph-smth} we only use the \PBZPWP signal and we vary the background. This is performed in the \MZP versus $\GZP/\MZP$ plane. The limits in Figure~\ref{fig:ph-smth} (top left), where the background is set to the data only for measurements where we have simulated predictions, are similar but slightly weaker than in Figure~\ref{fig:ph-smth} (top right), where the background is simulated at NLO accuracy. This is due to the fact that the SM predictions agree reasonably with the data, but overshoot it in some regions of the phase space. Furthermore, in Figure~\ref{fig:ph-smth} (bottom left), where the background is set to be equal to the NNLO predictions where available and to NLO elsewhere, the limits are slightly stronger than the ones in the top right panel of Figure~\ref{fig:ph-smth}. The reason is that at NNLO the scale uncertainties are reduced. Finally, in the bottom right part of Figure~\ref{fig:ph-smth}, we show the expected limits using the NNLO and NLO theory predictions.

The numerical results of Figure~\ref{fig:ph-smth} are summarised in Table~\ref{tab:results} together with the most stringent limits coming from a direct resonance search~\cite{CMS}. As explained above, we see that the limits are stronger with increasing precision. We also see that with \CONTUR one can explore areas of the phase space that are not accessible in direct searches. However, in the TC case, we see that the limits from the direct search are stronger than the ones obtained with \CONTUR. One reason is the fact that Run 2 measurements with full luminosity where not yet added to \RIVET. Another reason could be the difference in the binning between searches and measurements. While searches allow for empty bins when drawing the exclusions, measurements require at least some events to be in each bin.

\begin{table}
\caption{  -- Exclusion limits on \MZP obtained in this analysis. The last column shows the most stringent limits coming from a direct CMS search.}
\label{tab:results}
\begin{center}
\begin{tabular}{|c| c| c| c| c|} 
\hline
 \multicolumn{5}{|c|}{Excluded $M_{Z'}$ [Tev]}\\
 \hline
 $\Gamma_{Z'}/\MZP$ [\%] & Data as bgd. & NLO as bgd. & NNLO as bgd. & CMS \\ [0.5ex] 
 \hline
 1 & 2.29 & 2.35 & 2.50 & 3.80 \\ 
 \hline
 10 & 3.17 & 3.22 & 3.55 & 5.25 \\
 \hline
 30 & 4.01 & 4.04 & 4.53 & 6.65 \\
 \hline
 50 & 4.54 & 4.61 & 5.19 & -\\ 
\hline
\end{tabular}
\end{center}
\end{table}

\section*{Summary}

We performed an analysis using the different modes of \CONTUR. We validated the previous approach employed with \CONTUR (Data = SM) and used, for the first time, higher order calculations for both the signal (NLO) and the background (up to NNLO).
This analysis shed light on both the pros and cons of a \CONTUR
like approach. A wide range of BSM models can be rapidly checked using this toolkit and exclusion limits can be derived in regions of the parameter space that were not studied before. In the TC model case, the direct searches were more efficient than our analysis. However, for models with more free parameters and more complex phenomenology, or even for unexplored models, the \CONTUR approach is expected to be much more competitive.

\section*{Acknowledgments}

This work was supported by the National Science
Centre, Poland, under research grant \\ 2017/26/E/ST2/00135, the BMBF under 05H18PMCC1, the IN2P3 project “Théorie – BSMG”, and the European Union’s Horizon 2020 research and innovation programme as part of the Marie Sk lodowska-Curie Innovative Training Network MCnetITN3 (grant agreement no.~722104).


\section*{References}

\begin{thebibliography}{99}

\bibitem{JMB}J. M. Butterworth {\it et al}, \Journal{JHEP}{03}{078}{2017}  [1606.05296].

\bibitem{Buckley:2021neu}A. Buckley {\it et al}, \Journal{SciPost Phys. Core}{4}{013}{2021} [2102.04377].

\bibitem{rivet}C. Bierlich {\it et al}, \Journal{SciPost Phys.}{8}{026}{2020}  [1912.05451].

\bibitem{lepTC}R. M. Harris {\it et al}, \Journal{Eur. Phys. J. C}{72}{2072}{2012} 
[1112.4928].

\bibitem{ConturTC}M. M. Altakach {\it et al}, [arXiv:2111.15406 [hep-ph]].

\bibitem{altakach:1}M. M. Altakach {\it et al}, \Journal{Phys. Rev .D}{103}{115026}{2021} [2012.14855].

\bibitem{hvq}S. Frixione {\it et al}, \Journal{JHEP}{09}{126}{2007} [0707.3088].

\bibitem{NNLO}J. Mazzitelli {\it et al}, \Journal{Phys. Rev. Lett.}{127}{062001}{2021} [2012.14267].

\bibitem{dijet}S. Alioli {\it et al}, \Journal{JHEP}{04}{081}{2011} [1012.3380].

\bibitem{ATLAS}ATLAS Collaboration, \Journal{JHEP}{05}{195}{2018} [1711.02692].

\bibitem{CMS}CMS Collaboration, \Journal{JHEP}{04}{031}{2019} [1810.05905].





\end{thebibliography}
\end{document}